\documentclass[preprint]{emulateapj}

\usepackage{epsfig}
\usepackage{portland}

\slugcomment{To be Published in {\it{The Astronomical Journal}}}
\shortauthors{Hill \& Zaritsky}
\shorttitle{Structural Parameters of SMC Clusters}

\begin{document}

\title{The Star Clusters of the Small Magellanic Cloud: \break Structural Parameters}

\author{Andrew Hill\footnote{Currently at the Department of Astronomy \& Astrophysics, University of Chicago, 5640 S. Ellis Ave, Chicago, IL, 60637}  \  \& Dennis Zaritsky}

\affil{Steward Observatory, University of Arizona, 933
   North Cherry Avenue, Tuscon, AZ 85721, USA}
\email {arhill@uchicago.edu, dzaritsky@as.arizona.edu}

\begin{abstract}             
We present structural parameters for 204 stellar clusters in the Small Magellanic Cloud derived from fitting King and Elson, Fall, \& Freeman model profiles to the V-band surface brightness profiles as measured from the Magellanic Clouds Photometric Survey images.  Both King
and EFF profiles are satisfactory fits to the majority of the profiles although King profiles 
are generally slightly superior to the softened power-law profiles of Elson, Fall, and Freeman and
provide statistically acceptable fits to $\sim$90\% of the sample. We find no
correlation between the preferred model and cluster age.
The only systematic deviation in the surface brightness profiles that we identify is a lack of a central concentration in a subsample of clusters, which we designate as ``ring" clusters.  In agreement with previous studies, we find that the clusters in the SMC are significantly more elliptical than those in the Milky Way. However, given the mean age difference and the rapid destruction of these systems, the comparison between SMC and MW should not directly be interpreted as either a difference in the initial cluster properties or their subsequent evolution.  
We find that cluster ellipticity correlates with cluster mass more strongly than with cluster age. We identify several other correlations (central surface brightness vs. local background density, core radius vs.  tidal force, size vs. distance) that can be used to
constrain models of cluster evolution in the SMC.                             

\end{abstract}

\section{Introduction}

The study of the structure and evolution of stellar clusters has been reinvigorated
by the discovery of young star clusters in ongoing galaxy mergers \citep{holtzman92, whitmore} and
poststarburst systems \citep{yang}.
The desire to understand the formation and evolution of these objects, and to use these
objects to provide a measure of the time and efficiency of star formation during mergers,
provides yet another connection between studies of distant and Local Group galaxies.
The Magellanic Clouds contain numerous young clusters that can be studied in 
detail because of their proximity and can guide our understanding of cluster evolution. A
first step in the study of the cluster population is the measurement of such
physical characteristics 
as the surface brightness profiles. We present surface 
brightness profiles and model fits for 
204 clusters in the Small Magellanic Cloud.

Our understanding of the surface brightness profiles of star clusters is dominated by two particular 
models. The first generically successful set of models was empirically derived to 
reproduce the surface brightness profiles of Galactic globular clusters \citep{k62}. These have become known as King profiles and were later derived from tidally limited models of isothermal spheres \citep{k66}.  Consequently they are expected to accurately describe bound stellar systems, if they have isothermal and isotropic stellar distribution functions, stars of the same mass, and
reside within a tidal field exerted by another object.  Extensions to non-isotropic systems \citep{michie}, rotating systems \citep{pt,wilson}, anisotropic and rotating systems \citep{lg}, and multimass models \citep{daf, gg} exist.  Only clusters in the nearest galaxies, including the Milky Way of course, are sufficiently well-resolved to test whether King models, and the corresponding extensions, are proper descriptions. In general, the basic King model is sufficient to explain the observed profiles of Galactic globular clusters, typically a surface density or surface brightness profile, although the extensions have been tested in a few clusters with more detailed observations (for example, using radial velocities in M 13 \citep{lgg}). As such, they can be viewed 
either as highly successful fitting formulae to the surface brightness profiles 
or as self-consistent dynamical models. It is incorrect,
however, to infer the full dynamical properties of the King model solely on the basis of
a successful fit to the surface density profile. 

The second generically successful set of models was empirically derived to reproduce
the surface brightness profiles of young 
($<$ 300 Myr) clusters in the Magellanic Clouds \citep{e87, e91}, hereafter EFF profiles. These
models have also been successfully 
applied to young clusters in external galaxies \citep{whit}.
The lack of the developed tidal cutoff is associated with 
the cluster's dynamical youth, although 
profiles of some old Galactic clusters show similar features at low surface brightness
levels that are referred to as extra-tidal stars \citep{grill}. When, or even if, a cluster might evolve
from an EFF profile to a King profile is unknown and may depend on its environment
as well as its individual characteristics. The lack of a well established tidal limit 
was confirmed for one young LMC cluster (NGC 1866) on the basis of stellar kinematics 
and more marginally suggested
for two others (NGC 2164 and NGC 2214; \cite{lup})

Whether one type of profile is more appropriate for all clusters in the Magellanic Clouds, 
or whether the appropriate profile depends on cluster age is also unknown because the
existing samples of high-quality profiles 
preferentially include young clusters and because they
are small (18 clusters in the LMC \citep{e91} out of an estimated population 
of several thousand \citep{bicalmc}, and none in the SMC)\footnote{\cite{mac03a,mac03b} have recently obtained high quality surface brightness profiles of 10 SMC and 53 LMC clusters with
the {\sl Hubble Space Telescope}, but they note that
their profiles do not extend sufficiently far in radius to distinguish between King and EFF
models \citep{mac03a}.}.
Previous cluster catalogs have identified thousands of clusters in the Large and Small Magellanic Clouds (LMC and SMC; \cite{bicalmc,ogle,bica}) over a large range of ages \citep{van81}.
The combination of structural parameters and ages is critical in unraveling the evolution of these clusters, and by extension the processes that drive cluster evolution in general. Many
issues, such as the relationship between core radius and age \citep{e89} or ellipticity and
age \citep{FF} are independent of whether an EFF or King profile is most appropriate at large
radius.

Even at the distance of the Magellanic Clouds it can often be difficult to resolve individual cluster stars, and hence both the quality of the data and the details of the model fitting become important.
Studies of cluster structure fall into two camps. Either they have used high quality data for a few
clusters, such as the study  by \cite{mac03a,mac03b} who use {\sl Hubble Space Telescope} images 
for 10 clusters in the SMC, or they have used photographic plates, such as the study by
\cite{kon82},  which achieve a wider field of view but are unable to resolve the inner structure of any cluster.  We pursue a study in the middle ground 
by adopting modern data and algorithms to measure and fit the cluster luminosity profiles, but use 
ground based data for which determining the structure inside of a few arcsec is limited. We present the
necessary basic structural parameters for 204 SMC clusters. 

This study has two principal purposes. First,
we aim  to determine if clusters, particularly those currently dissolving, are still well described by a King profile, and if the older clusters are well described by an EFF profile. Second, aside
from issues related to the outer profiles, we present measurements such
as core radius and ellipticity that are potentially related to cluster evolution and independent
of the detailed shape of the outer surface brightness profile.
In addition to the results discussed here, 
these measurements 
provide a database from which future, more detailed studies can select 
targets and form a basis for comparison with other measurements.
For example, cluster ages are presented by \cite{rz} and a subsequent study will attempt to place both sets of data within a consistent evolutionary model.  In \S2 we describe our cluster catalog, in \S3 we present our methodology for fitting King and EFF profiles and compare our results to previous studies, in \S4 we develop some inferences from the 
results, and we summarize in \S5.

\section{Data and Cluster Catalog}
	
	The data come from the Magellanic Clouds Photometric Survey (MCPS)  described by \cite{z97} and \cite{z02}. In summary, we have obtained U,B,V, and I drift-scan images of the central 4.5$^\circ$ by 4$^\circ$ area of the SMC using the the Las Campanas Swope (1m) telescope and the Great Circle Camera (\cite{z96}). The stellar photometric catalog is provided by \cite{z02} and we utilize it and the original images in this study.  The effective exposure time is between 4 and 5 min for the SMC scans and the pixel scale is 0.7 arcsec pixel$^{-1}$. Typical seeing is $\sim$ 1.5 arcsec and scans with seeing worse that $\sim$2.5 arcsec were rejected. The final catalog contains over 5 million stars with at least both $B$ and $V$ measured magnitudes. The astrometry for individual stars has a standard deviation of  0.3 arcsec (dervied both from internal errors and comparisons with a variety of external studies, see \cite{z02}).

	Although extensive cluster catalogs already exist for the SMC, we performed our own cluster search to examine the completeness of those catalogs over the entire MCPS survey area. The approach we use is similar to that employed for a portion of the LMC by \cite{z97} and for the central ridge of the SMC by \cite{ogle}. Specifically, we construct a stellar density image using the stellar photometric catalog. Within 10 arcsec pixels, we count the number of stars with $V < 20.5$. This particular magnitude limit is chosen to minimize spurious detections and spatial variations due to differing levels of completeness within and across different scans. The resulting stellar density image is then median filtered with a box of size 10 such pixels (100 arcsec), and the filtered image is subtracted from the unfiltered image to remove large-scale stellar density fluctuations.  
We use SExtractor \citep{sextractor} to identify peaks in the residual image and set the parameters generously so that many spurious peaks are found. 
	
	Our need in this particular study is for unambiguous clusters for which structural parameters can be measured. Most of the effort in building a cluster catalog goes into recovering and confirming the most marginal objects.  As such, some of the clusters published in previous studies are quite dubious and various authors disagree on the validity of these poor clusters (for example, see the cross-comparison of the OGLE catalog \citep{ogle} with the previous catalog by \cite{bica}). Because our aim is to proceed with further detailed study of clusters, we focus on defining a set of robust clusters. From a cross-comparison of our catalog with published catalogs, and then visually inspecting the images of candidates, we settle on the list of 204 clusters presented in Table 
\ref{tab:Names and Positions}.  Of the rejected clusters from our catalog, many match identifications in other catalogs but are either visually unconvincing,  too poor to offer much chance of further investigation, or embedded in strong nebular emission. Therefore, exclusion from our list does not necessarily imply that we reject that candidate cluster as a real cluster. The added criteria that we have imposed will preferentially exclude young clusters within emission regions, those embedded in the denser parts of the SMC, and those with a relatively low central surface brightness. This sample is not complete, nor necessarily representative of the overall population, but the biases are qualitatively understood and the sample is still far larger than any for which detailed structural parameter fits are available.

\begin{deluxetable}{cccr}
\tabletypesize{\tiny}
\tablecaption{Cluster Designation and Position}
\tablewidth{0pt}

\tablehead{
\colhead{Cluster} &
\colhead{RA} &
\colhead{DEC} &
\colhead{Alternate} \\
&&&{Designations}
}

\startdata

1	& $0^{h}  24^{m}  18.67^{s}$	& $-73^{\circ}  59^{\prime}  35.8^{\prime \prime}$	& \nodata	\\
2	& $0^{h}  24^{m}  43.16^{s}$	& $-73^{\circ}  45^{\prime}  11.7^{\prime \prime}$	& K5, L7, ESO28-18 	\\
3	& $0^{h}  25^{m}  26.60^{s}$	& $-74^{\circ}  04^{\prime}  29.7^{\prime \prime}$	& K6, L9, ESO28-20		\\
4	& $0^{h}  27^{m}  45.17^{s}$	& $-72^{\circ}  46^{\prime}  52.5^{\prime \prime}$	& K7, L11, ESO28-22	\\
5	& $0^{h}  28^{m}  02.14^{s}$ 	& $-73^{\circ}  18^{\prime}  13.6^{\prime \prime}$	& K8, L12		\\
6	& $0^{h}  29^{m}  55.22^{s}$	& $-73^{\circ}  41^{\prime}  57.1^{\prime \prime}$	& HW3	\\
7	& $0^{h}  30^{m}  00.27^{s}$ 	& $-73^{\circ}  22^{\prime}  40.7^{\prime \prime}$	& K9, L13		\\
8	& $0^{h}  31^{m}  01.34^{s}$	& $-72^{\circ}  20^{\prime}  29.0^{\prime \prime}$	& HW5	\\
9	& $0^{h}  32^{m}  41.02^{s}$	& $-72^{\circ}  34^{\prime}  50.1^{\prime \prime}$	& L14	\\
10	& $0^{h}  32^{m}  56.26^{s}$	& $-73^{\circ}  06^{\prime}  56.6^{\prime \prime}$	& NGC152, K10, \\
&&&L15, ESO28-24\\

\enddata
\tablecomments{The complete version of this table is in the electronic edition of
the Journal.  The printed edition contains only a sample.}
\label{tab:Names and Positions}
\end{deluxetable}

         For the crossidentification presented in Table \ref{tab:Names and Positions}, we identify the closest cluster in projection from the \cite{bica} catalog which contains clusters, association, and some other sources such as planetary nebulae.   For matches, the coordinates typically differ by a few arcsec, well within the quoted general uncertainty of 10 arcsec (\cite{bica}). 

\section{Profile Fitting}

	\subsection{Methodology}
	\label{sec:Methodology}
	
\begin{figure}[htbp]
\begin{center}
\plotone{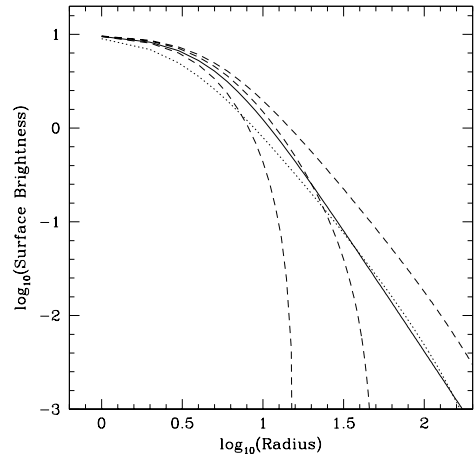}
\caption{A comparison of King and EFF profiles (arbitrary units). The short dashed lines represent
King models with $c$ = 0.52, 1, 2.1 and an identical core radius (5), 
with increasing $c$ going from left to right. The solid line represents an EFF profile with the 
same value for $a$ and a value of $\gamma$ typical for young clusters, 2.6 \citep{e87}.
The dotted line represents a King model with $c = 2.1$ and a core radius adjusted to 
result in an outer profile that agrees well with that of the EFF model. }
\label{fig:profile comparison}
\end{center}
\end{figure}

	Although we fit both King and EFF profiles to our clusters, we place greater emphasis on the King profiles for several reasons.  First, the full set of King models include profiles with extended wings.  The largest concentration models we fit resemble a power law profile with an effective index similar to the typical EFF profile (see Figure \ref{fig:profile comparison}).  Second, previous studies have only demonstrated that the youngest LMC clusters, those with age $<300$ Myr, are better fit by an EFF profile, while over half of our cluster sample is estimated to be older \citep{rz}.  Finally, focusing on a single family of models allows for a more straightforward comparison across the sample.  Our decision to emphasize the King profile is certainly open to critique, particularly if the data were to favor the EFF profile.  Our results
indicate that this is not the case. If the surface brightness profiles of young clusters are better fit by softened power law profiles, we should find these clusters more successfully fit by increasingly concentrated King models.  However, our analysis shows King models are not only statistically satisfactory fits for $\sim$ 90\% of the clusters, but also are generally superior fits than EFF profiles for the bulk of the sample (see \S3.2).

	The usual approach of fitting a model profile to the surface density of a cluster is complicated by the incompleteness of our stellar catalog that results from crowding in the centers of many clusters.  We overcome this difficulty by fitting models to the surface brightness profile rather than attempting the difficult and inherently uncertain task of defining and applying an incompleteness correction.  We choose to work with the $V$-band images because they are the deepest of our set, and because $V$ is more sensitive to the underlying stellar populations than either $U$ or $B$ while being less sensitive to foreground contaminating stars than $I$.  Furthermore, use of $V$ allows us to compare directly to previous studies \citep{kon82, kon83, kon86, mac03b} without concern for possible color gradients.
	
	Obtaining an accurate determination of each cluster's center is the first step in constructing the surface brightness profile.  We use the approximate coordinates given in our catalog to extract a 350$^{\prime\prime} \times 350^{\prime\prime}$ image of the cluster.  A precise center is determined by calculating the luminosity-weighted centroid.  To avoid spurious results due to contamination by other clusters or bright foreground stars, we constrain the calculation to pixels within 20$^{\prime\prime}$ of the approximate center and exclude pixel values that are 3000 counts above the mode of the entire frame (typical mode values are between 100 and 200 counts).  We iterate four times, with each successive run using the previous luminosity-weighted center.  The derived centers are stable.  For some irregular clusters, we adjust the center to match a visually identified center that corresponds to central condensation of stars, rather than a broader luminosity centroid.  This adjustment is typically not more than a few arcsec.
	
	As has been previously noted \citep{FF}, many SMC clusters are elliptical.  The nature of this ellipticity, whether it is fundamentally connected to age or luminosity, is debated \citep{e91, van84}.  Despite the conclusion from \cite{e91} that the asphericity primarily results from unrelaxed stellar concentrations rather than any true underlying elliptical light distribution, we opt to effectively use elliptical isophotes in determining our surface brightness profile.  Because we model the surface brightness of the cluster, subconcentrations should be considered both in determining whether a profile is a good fit and whether elliptical isophotes are necessary.
	
	The ellipticity and position angle of each cluster are determined by computing the second moments of the cluster luminosity distribution (see \cite{banks} for a similar treatment of a small sample of LMC clusters). We restrict the computation to pixels with fewer than 5000 counts above the mode, and to radii between 3$^{\prime\prime}$ and 0.7$r_t$, where $r_t$ is the fitted tidal radius, or 30$^{\prime\prime}$ on the first iteration of this procedure. These criteria are selected with the intention of minimizing the effect of background and foreground contamination and ignoring the seeing-smoothed inner radii.  The ratio of the eigenvalues of the matrix formed from the second moments of the luminosity distribution is the projected ratio of the semi-minor, $b$, to semi-major, $a$, axis of the cluster, while the eigenvector matrix is the rotation matrix of the cluster.  Using these results, we rotate the image so that the major axis lies along the horizontal axis and scale the image along the vertical axis by $a/b$.
	
	We compute the surface brightness profiles using circular annuli of  2$^{\prime\prime}$ width and adopt an uncertainty that is given by $\sigma$ of the mean of the pixel values within that annulus. We exclude pixels with more than 12000 counts above the mean for the annulus to eliminate contamination from the brightest stars.  We choose different pixel thresholds for our calculation of the center, ellipticity, and surface brightness because of different sensitivities to non-cluster objects. Surface brightness profiles are measured out to 200$^{\prime\prime}$ and include a substantial sky region for all clusters.

\begin{figure}[htbp]
\begin{center}
\plotone{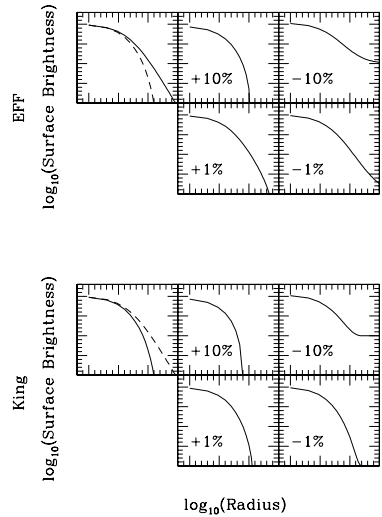}
\caption{A comparison of the effect of background subtraction errors on both representative
King and EFF profiles. Upper set of panels illustrate the effect of oversubtracting ($+$) 
and undersubtracting ($-$) the background by a percentage of the central brightness
of the cluster on EFF profiles (the original profile is represented by the solid line in the 
upper left panel). The dashed line represents the comparison King profile in the
upper panel and the comparison EFF profile in the lower panel. The lower panels
are analogous except for illustrating the effect of background errors on a King profile.}
\label{fig:profile background}
\end{center}
\end{figure}

\begin{figure*}[htbp]
\begin{center}
\plotone{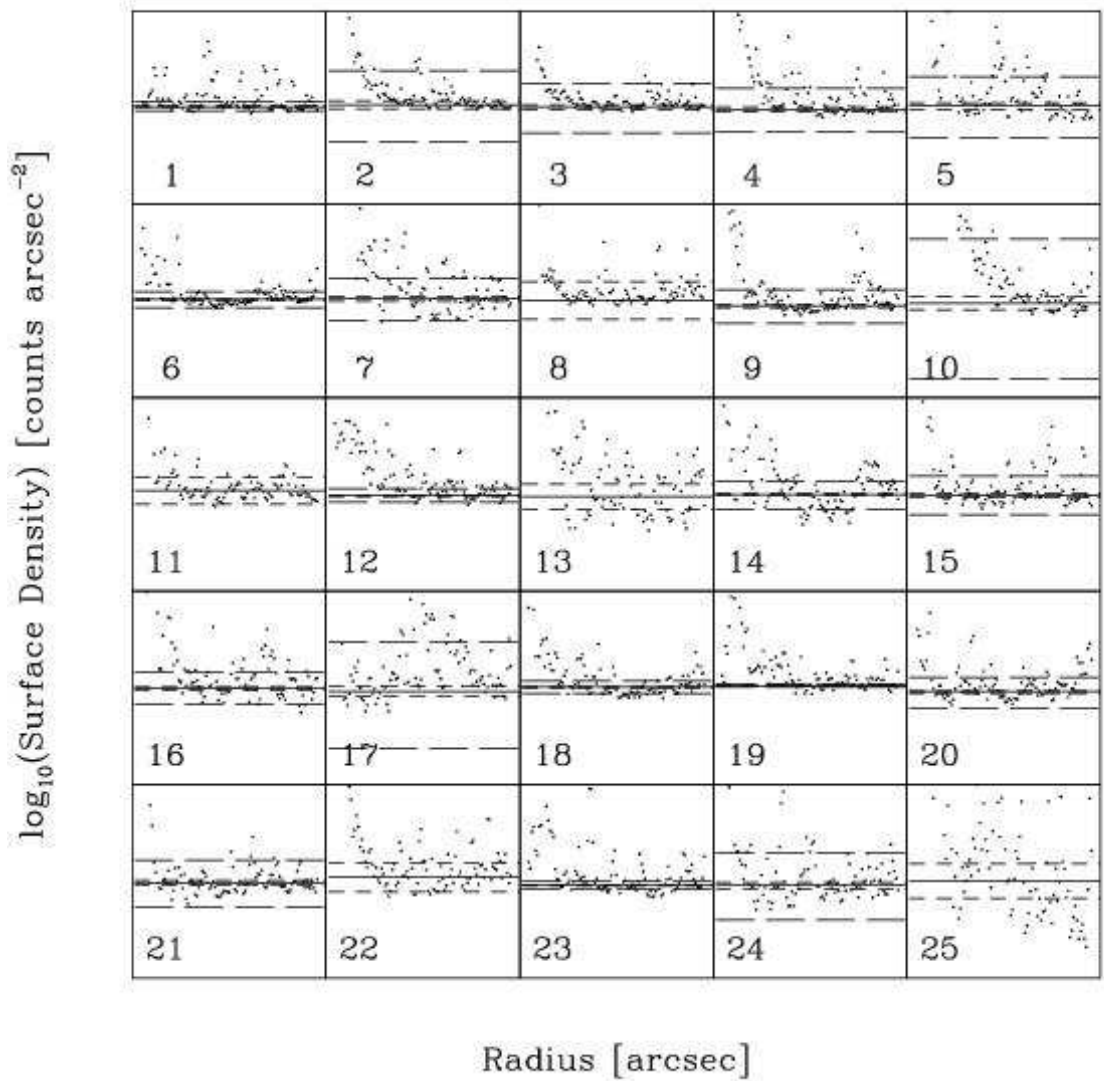}
\caption{Background fits for the first 25 SMC clusters (numbers in lower right
are the cluster designation).
The range of horizontal axis is from 0 to 200 arcsec for all clusters. The
vertical scale is unique for each cluster and corresponds to $\pm$ 10\% of the selected
background
value. The sky used in our fitting is shown by the solid line. The short dashed line 
mark the range of error corresponding to 1\% of the central surface brightness of the cluster
and the long dashed lines correspond to 10\%.
We show a representative sample here and present all the clusters in the electronic 
version of the Figure.}
\label{fig:backgrounds}
\end{center}
\end{figure*}

	The determination of the sky level is key in assessing the nature of the outer regions of the surface brightness profiles of clusters. As shown in Figure \ref{fig:profile background}, an error equivalent to 10\% of the central surface brightness of the cluster can convert an intrinsic EFF profile into an apparently tidally truncated profile or an intrinsic King profile into an apparent one with no tidal radius. Errors in the background that are less than 1\% of the cluster's central surface brightness do not qualitatively affect the nature of the outer radial profile.  It is important to note that this is a statistical uncertainty that also affects, but was ignored by, previous studies.  We measure the background using an outer annulus that typically extends over the last quarter of the radial range of the extracted images ($r > 150^{\prime\prime}$), although we adjust the boundaries of the sky region interactively for large clusters or for those with contaminating stars.  Because young stars are spatially correlated \citep{hz}, going to larger and larger radius to measure the background is not an appropriate solution, particularly for the younger clusters where the spatial correlations are the strongest.   Moreover, there is no reason to expect the background at large radii from the cluster to be the same as the background in which the cluster resides, especially if there are many young stars in the region.  Unfortunately, adopting background annuli near the clusters can often result in background measurements that are affected by local fluctuations. The determination of the background is therefore an inherently poorly solved problem. We opt against including the background as a free parameter in the model fitting described below because we find that including radial bins in which the background dominates allowed us to improve the statistical measure of the fit, although it often introduces systematic errors in the fitted cluster profile.  This problem arises because the photometric properties of the area around the cluster are not necessarily identical to those underlying the cluster, in particular the background is not uniform. A statistical fitting method attempts to fit the background at the expense of fitting the cluster and results in fits that are visually inferior in many cases.  Including only radial bins in which the cluster dominates results in fitting degeneracies due to the lack of a strong direct constraint on the background level.  Instead, we determine the background independently from a sky annulus at moderate radius, which is similar to the procedure adopted by EFF.  The inner boundary of our sky region is typically about 8 times larger than the radius that includes 90\% of the cluster light.  Our fits to the background, and the comparison benchmarks of 1 and 10\% errors,  are shown in Figure \ref{fig:backgrounds}. The Figure illustrates that there are some clusters for which the scatter in the background exceeds the 10\% benchmark and hence sky uncertainties play a significant role, while there are others for which the scatter is less than the 1\% benchmark and hence sky uncertainties are sufficiently small that the outer cluster profile is well measured.

	We fit King and EFF profiles  to each of our clusters via chi-squared minimization.
The standard King model is described by a combination of three parameters \citep{k66}.  We choose to define each profile by its central surface brightness, $\Sigma_0$, core radius, $r_c$, and concentration, $c$. Other common parameters, such as tidal radius, can be derived from this set (ie. $c \equiv log_{10}({r_t / r_c})$).  In addition to these three parameters, our profiles include the background surface brightness discussed above.  We explore a parameter space that ranges in $\Sigma_0$ from 0.25 to 1.75 times the mean $\Sigma$ measured within the central 4$^{\prime\prime}$ of the cluster in increments of 0.025, in $r_c$ from 1$^{\prime\prime}$ to 101$^{\prime\prime}$ in increments of 0.5$^{\prime\prime}$, and in $c$ from 0.001 to 2.1 in increments of 0.01.  The concentration range corresponds to the range over which King's original empirical models match his theoretical models as found by \cite{w85}.  The fitting is iterated once.  Error bounds on each best fit parameter corresponds to the range of that parameter among models that cannot be rejected with more than 67\% confidence. 

\begin{figure*}[htbp]
\begin{center}
\plotone{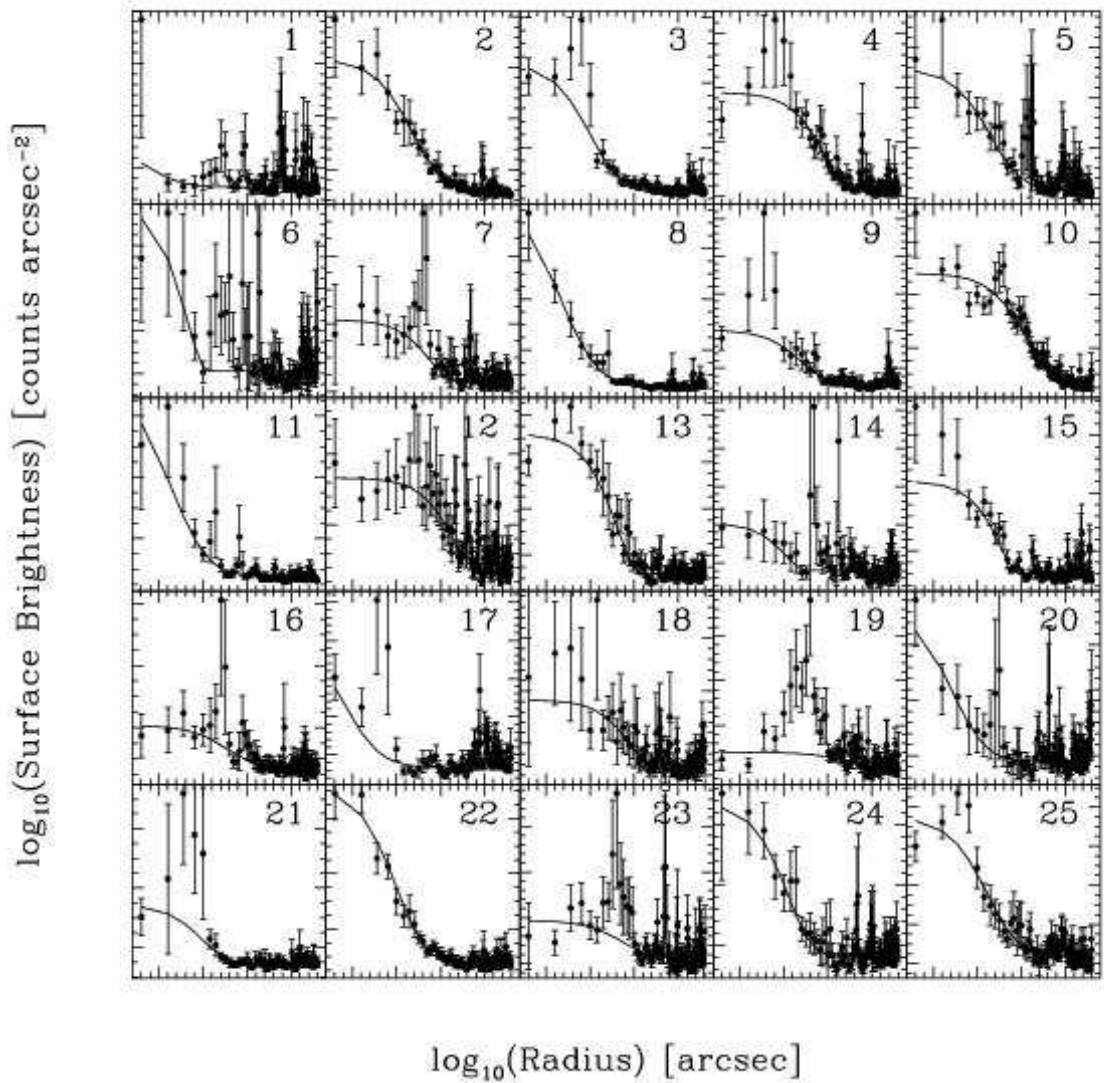}
\caption{$V$-band surface brightness profiles of the first 25 SMC clusters and profile fits. 
These profiles include the background flux.
The range of horizontal axis is from 0 to 200 arcsec for all clusters. The
vertical scale is unique for each cluster. The central surface brightness of each cluster
fit is given in Table \ref{tab:Structural Parameters}. The best fit King profile
is shown by the solid line. The numeral in the upper right of each panel
provides the cluster identification. 
We show a representative sample here and present all the clusters in the electronic 
version of the Figure.}
\label{fig:Sample Profiles}
\end{center}
\end{figure*}
\begin{figure*}[htbp]
\begin{center}
\plotone{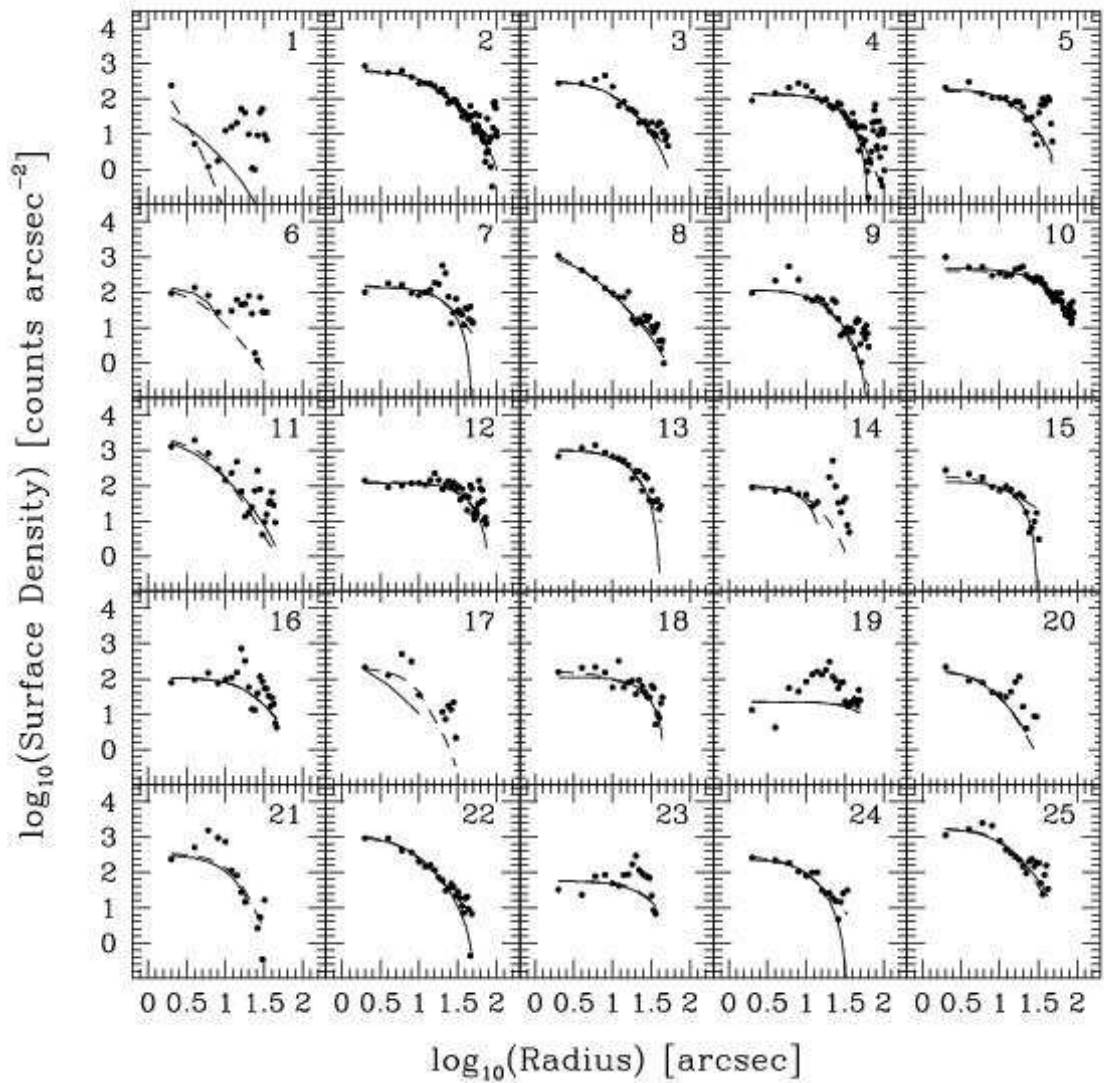}
\caption{The $V$-band cluster profiles, after background subtraction, are compared to 
both the best fitting King (solid line) and EFF profiles (dashed line). The 
radial plots are truncated at the innermost radius for which the counts minus the background are less than zero
to avoid only plotting values for which the logarithm exists. This omission results
in some profiles that appear to be systematically poor fits because the fit includes
data that are omitted from the Figure.  Error bars have been omitted for clarity.
All panels are identically scaled and the numeral in
the upper right corner provides the cluster identification. All 
cluster profiles are available in the electronic 
version of this Figure. }
\label{fig:Sky Sub}
\end{center}
\end{figure*}

As mentioned above, \cite{e87} noted that the surface brightness profile, $\mu(r)$, of young clusters appear to be better described by a core plus power law profile, $\mu(r) \propto (1 + r^2/a^2)^{-\gamma/2}$, where $a$ and $\gamma$ are free parameters.  We fit the EFF models our data in the same manner as we fit the King profiles, over the same radial range, by varying the central density ($\Sigma_0$), the
scale length ($a$), and the power-law index ($\gamma$). Note that $a$ is the scale length, rather
than the core radius, which can be calculated using $r_c = a\sqrt{2^{2/\gamma}-1}$.
We explore a parameter space that ranges in $\Sigma_0$ from 0.25 to 1.75 times the mean $\Sigma$ measured within the central 4\arcsec of the cluster in increments
of 0.025, in $a$ in from 1$\arcsec$ to 101 $\arcsec$ in increments of 0.5\arcsec, and in 
$\gamma$ from 1.5 to 7.0 in increments of 0.05. We present the best fit King profiles in Figure \ref{fig:Sample Profiles} and both King and EFF profiles in Figure \ref{fig:Sky Sub} overlayed on the data. The key difference between the two Figures is that Figure \ref{fig:Sample Profiles} shows the un-background subtracted data so that all of the data could be plotted on the logarithmic scale. Note however, the profiles plotted in this manner do not resemble the standard King models, although they are. In Figure \ref{fig:Sky Sub} we again plot the background-subtracted profiles, but here we stop plotting data and models once a radial bin has a negative background subtracted value because the logarithm is undefined. Plotting only data greater than the background, i.e. those with positive background subtracted values, results in a misleading profile.  The parameter values for the fits are presented in Table \ref{tab:Structural Parameters}.  The $1\sigma$ confidence intervals are defined
using $\chi^2$, and are therefore undefined in the cases where even the best fit model has $\chi^2$
greater than the corresponding value for the available degrees of freedom.

	\subsection{Results}
	\label{sec:Results}

	Of the 204 clusters in our catalog, only 44 are poorly described by a King profile with greater than 67\% confidence.  Given the size of our sample however, we expect from statistical considerations that a substantial number of clusters that are truly well described by a King profile will be rejected at this level. We calculate that only 19 clusters are fit sufficiently poorly that for each individual cluster the King profile can be rejected as an accurate description of their surface brightness profile (this limit corresponds to rejecting the profile with 99.5\% confidence). Because both the King and EFF profiles can fit clusters with extended radial surface brightness profiles, the profile itself cannot be used
to infer extratidal stars (i.e. stars beyond the cluster's Roche lobe). While the extended wings in 
the EFF profile are generally interpreted to arise from that type of mass loss, the extended
wings in a high-concentration King model are bound to the system. Determining which
physical interpretation is correct will require either direct dynamical measurements, such
as a measurement of the cluster mass and the velocity dispersion of stars, or indirect
evidence, such as a lack of old clusters with such wings, which would imply that there has
been dynamical evolution of such clusters. In this study, we reach conclusions
on the relative merits of the fitting functions, not on the dynamical state of the clusters.
As we have illustrated before (Figure \ref{fig:profile comparison}), describing 
a surface brightness with a King profile does not necessarily imply the presence of a 
dramatic tidal edge.

\begin{figure}[htbp]
\begin{center}
\plotone{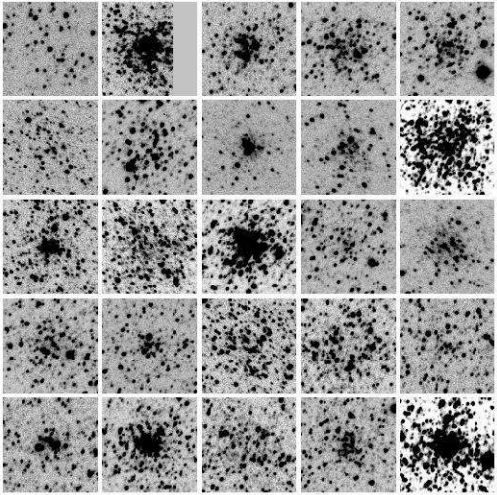}
\caption{$V$-band images of the first 25 SMC clusters. The angular size of 
each image corresponds to the central quarter of the area for which the profiles 
in Figure \ref{fig:Sample Profiles} are derived.
Images of all the clusters in the four filters are available in \cite{rz}.}
\label{fig:Sample Images}
\end{center}
\end{figure}

\begin{figure}[htbp]
\begin{center}
\plotone{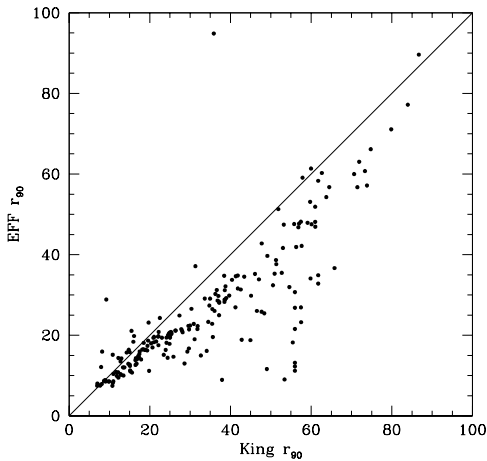}
\caption{Comparison of $r_{90}$ values obtained with the King and EFF models.
The solid line is the 1:1 line.}
\label{fig:r90 comparison}
\end{center}
\end{figure}

\begin{figure}[htbp]
\begin{center}
\plotone{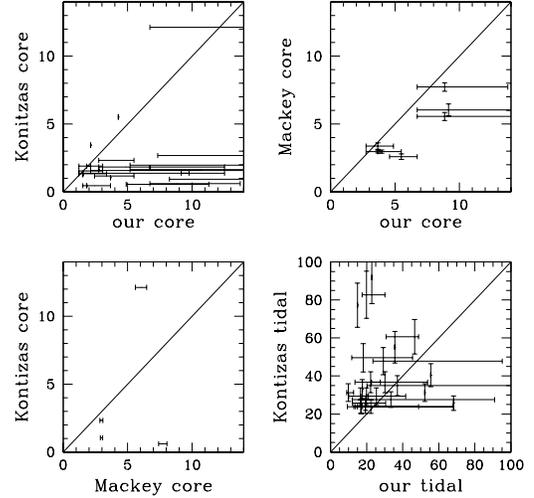}
\caption{Comparison of our measured core and tidal radii to previously published values.  Our core radii are compared to those presented in Kontizas et al (1982,1983,1986) in the upper left panel and those measured by Mackey \& Gilmore (2003b) in the upper right.  The core radii of Kontizas et al and Mackey \& Gilmore are compared at bottom left.  The bottom right panel features a comparison on our tidal radii to those published by Kontizas et al.  Clusters at (135,26) in the upper left and (55,207) in the bottom right were excluded for clarity.}
\label{fig:all_comp}
\end{center}
\end{figure}

	We find that our determination of the tidal radius is highly uncertain.  This uncertainty arises from the strong coupling of the estimated $r_t$ to the determination of the background.   Similarly, the uncertainty on the concentration, because it depends on $r_t$, is too large to be meaningful.  Because we desire a robust measure of cluster size, we pursue an alternative definition of size determined by the enclosure of a significant fraction of the total cluster light, rather than all of the light.  We choose to tabulate the radius inside of which 90\% of the total cluster luminosity of the {\sl fitted profile} is contained, $r_{90}$. 
Even though $r_{90}$, by definition, still contains the bulk of the cluster light, the associated uncertainty is typically half that of $r_t$. We avoid going to even smaller radii because then connection to the cluster size becomes increasingly tenuous and dependent on the cluster profile. Because of the ambiguity of which model profile to use, one may wonder whether defining $r_{90}$ using one profile or the other leads to large differences in the calculated value of $r_{90}$.  We show in Figure \ref{fig:r90 comparison} a comparison of the values of $r_{90}$ obtained using the best fit King or EFF model. We find that although there is some systematic difference between the two that can be as large as 25\%, the rankings of clusters according to size (and hence any correlation one might examine using $r_{90}$) is maintained to large degree regardless of the model one adopts.  The best fit parameters for each candidate, along with the 1$\sigma$ uncertainties, are presented in Table \ref{tab:Structural Parameters}.  The listed magnitude is calculated by integrating the fitted King profile.  The complete set of profiles, images, and structural parameters is presented in the online version of these Figures and Tables.

\begin{deluxetable*}{ccccccccccccrrccccc}
\tabletypesize{\tiny}
\tablecaption{Structural Parameters\tablenotemark{a}}
\tablewidth{0pt}
\tablehead{
\colhead{Cluster} &
\multicolumn{3}{c}{Core Radius [pc]} & \multicolumn{3}{c}{90\% Light Radius [pc]}
& \colhead{M$_V$} & \colhead{$\Sigma_{0,K}$ } & \colhead{$\epsilon$} & \colhead{$\chi^2_{\nu,K}$ } 
& \multicolumn{3}{c}{EFF Scalelength\tablenotemark{b} [pc]} & \multicolumn{3}{c}{$\gamma$} &
\colhead{$\Sigma_{0,E}$} &
\colhead{$\chi^{2}_{\nu,E}$} \\
 & 
\colhead{low}   & 
\colhead{best}   &
\colhead{high} &
\colhead{low} &
\colhead{best} & 
\colhead{high} &
&
\colhead{[cts/} &
&
&
\colhead{low} &
\colhead{best} &
\colhead{high} &
\colhead{low} &
\colhead{best} &
\colhead{high} &
\colhead{[cts/]}\\
&&&&&&&&arcsec$^{2}$]&&&&&&&&&arcsec$^2$]\\
}

\startdata

1&$0.61$&$0.61$&$0.92$&$11.58$&$15.56$&$17.34$&$18.7$&60&$0.30$&$0.99$&\nodata&$0.76$&\nodata&\nodata&$6.30$&\nodata&421&$1.24$\\
2&$2.44$&$3.97$&$7.33$&$12.17$&$18.06$&$20.84$&$12.9$&595&$0.17$&$0.77$&$1.07$&$3.35$&$8.37$&$1.50$&$2.20$&$4.65$&659&$0.70$\\
3&\nodata&$2.75$&\nodata&\nodata&$11.30$&\nodata&$14.5$&314&$0.02$&$1.25$&\nodata&$2.59$&\nodata&\nodata&$2.45$&\nodata&335&$1.93$\\
4&$5.19$&$10.69$&$18.94$&$11.54$&$12.62$&$16.69$&$13.8$&133&$0.18$&$0.81$&\nodata&$13.01$&\nodata&\nodata&$6.95$&\nodata&144&$1.50$\\
5&$1.53$&$3.97$&$12.52$&$6.43$&$11.90$&$19.20$&$14.8$&189&$0.28$&$0.68$&$1.22$&$8.52$&$10.96$&$1.50$&$6.90$&$7.00$&167&$0.73$\\
6&$0.61$&$3.36$&$14.97$&$10.39$&$18.31$&$24.30$&$16.9$&145&$0.37$&$0.79$&$0.61$&$1.37$&$12.94$&$1.50$&$2.75$&$7.00$&135&$0.82$\\
7&$9.16$&$9.16$&$9.77$&$9.57$&$9.57$&$9.62$&$14.5$&139&$0.56$&$1.10$&$3.51$&$6.55$&$11.87$&$1.85$&$3.40$&$7.00$&153&$1.08$\\
8&$0.92$&$0.92$&$1.53$&$7.08$&$11.78$&$14.98$&$14.7$&1259&$0.38$&$0.87$&$0.61$&$0.76$&$1.83$&$1.80$&$2.20$&$3.10$&1835&$0.87$\\
9&$2.75$&$4.89$&$12.52$&$6.99$&$11.24$&$12.81$&$14.8$&116&$0.37$&$0.70$&$3.20$&$8.68$&$10.50$&$2.25$&$7.00$&$7.00$&116&$0.87$\\
10&$6.72$&$9.16$&$14.36$&$19.82$&$22.40$&$24.11$&$12.1$&474&$0.23$&$0.79$&\nodata&$18.57$&\nodata&\nodata&$6.30$&\nodata&399&$1.35$\\
\enddata

\tablecomments{The complete version of this Table is in the electronic edition of
the Journal.  The printed edition contains only a sample.}
\tablenotetext{a}{The 1$\sigma$ confidence intervals are defined using $\chi^2$, and are
therefore undefined in the cases where even the best fit model has $\chi^2$
greater than the corresponding value for the available degrees of freedom.}
\tablenotetext{b}{ The core radius, when defined as the radius for which the surface
brightness drops to half its central value, can be calculated using $r_c = a\sqrt{2^{2/\gamma}-1}$.}
\label{tab:Structural Parameters}
\end{deluxetable*}

	Our measurements of structural parameters constitute the largest such survey of SMC cluster profiles, so the availability of comparison data is limited.  There exist only four published studies of SMC cluster profiles, three of which are by a single group of authors and should be considered a single survey.  \cite{kon82}, \cite{kon83}, and \cite{kon86}, hereafter collectively referred to as Kontizas et al., use photographic plates from the U.K. Schmidt and Anglo-Australian telescopes in Australia to measure 67 cluster profiles, while \cite{mac03b} present 10 profiles measured from images obtained with the {\sl Hubble Space Telescope}.  All 10 clusters observed by \cite{mac03b} appear in Kontizas et al., but most were observed using different filter passbands. Although we do not know whether the structural parameters are strongly dependent on color, a conservative comparison should be limited to parameters determined from images taken with similar filters. An examination of Kontizas et al.'s  observations in a variety of filters and emulsions do suggest that a cluster's profile may depend on the wavelength in which it is observed, although this issue is not specifically investigated.  We restrict our comparison to the smaller set of 28 clusters observed in a filter and emulsion combination closely corresponding to our $V$-band observations.  Our cluster sample includes 25 from this restricted set.
	
	Our measurements of $r_c$ and $r_t$ are compared to those made by Kontizas et al. and \cite{mac03b} in Figure \ref{fig:all_comp}.  Our core radii are systematically larger than those measured by Kontizas et al.  This result is consistent with Kontizas et al.'s prediction that their core radii are underestimated, a conclusion also reached by \cite{mac03b}.  Given that Kontizas et al. could not resolve the core in any of their clusters, it is not surprising that there is disagreement with both our measurements and those of \cite{mac03b}.  More illuminating is the comparison of $r_c$ measurements with those of \cite{mac03b}, who are able to resolve the inner regions of the clusters.  Here we find that our results are statistically consistent, although with some slight bias to overestimate $r_c$ by $\sim$ 30\%.  We suspect that our core radii are systematically larger because we are unable to resolve the inner cluster structure to the degree possible in the {\sl HST} observations. 
	
	Comparison of tidal radii is more uncertain both because there is a smaller set of data with which to compare to and the lack of modern techniques used in constructing smaller comparison set.  Although the \cite{mac03b} observations do not extend to large enough radii, Kontizas et al's do.   Figure \ref{fig:all_comp} reveals that our tidal measurements are consistent with, although $\sim$ 20\% systematically smaller than, those reported by Kontizas et al.  Given the lack of precision with which tidal radius can be fit (see \S\ref{sec:Results}), this difference is not surprising, especially given the significant difference in the quality of the data and analysis techniques between our study and that by Kontizas et al.  We conclude that our model fits are in rough accord with previous fits, but given the size 
of the scatter seen in Figure \ref{fig:all_comp} and our calculated uncertainties for
$r_t$ we do not include $r_t$ values in Table \ref{tab:Structural Parameters}.

\begin{figure}[tbp]
\begin{center}
\plotone{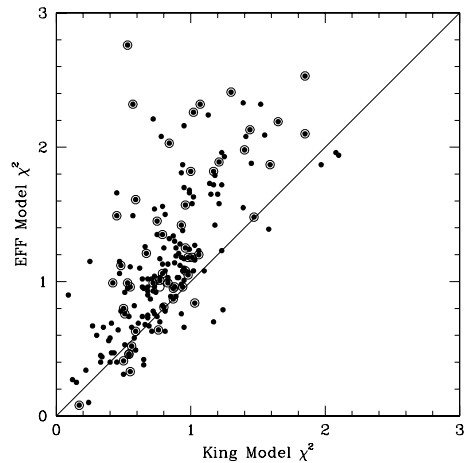}
\caption{The $\chi_\nu^2$ values for the best fit King and EFF profiles. The line is the
1:1 line. Only clusters for which the fits have $\chi_\nu^2 < 3$ are shown. Those clusters with
background values determined to better than 1\% of the central cluster surface 
brightness are plotted with larger open circles surrounding the central filled circle.}
\label{fig:chi2}
\end{center}
\end{figure}
	We now return to the issue of the appropriateness of the King model. There are two fundamental issues: 1) does the King profile provide an acceptable description of the surface brightness profile and 2) does the King model provide an acceptable description of the stellar dynamics. With our data, we are only in a position to address the former of these. Indeed, dynamical considerations would suggest, as noted above, that King models are unlikely to be the correct dynamical description of the younger clusters.  For the bulk of the clusters there are only subtle differences between the King and EFF profile that mostly occur near the outer edge of the clusters, in the region where the background uncertainties may dominate.  The fit in this region is not shown in Figure \ref{fig:Sky Sub} because we stop plotting once the background subtracted flux value is less than zero, and is hence undefined in the units plotted. 
However a comparison of the $\chi_{\nu}^2$ values, Figure \ref{fig:chi2}, demonstrates that  while for any individual cluster the $\Delta \chi_{\nu}^2$ is typically not large enough to rule out one model relative to the other, the King model systematically fits these cluster profiles better.  To address the importance of the background subtraction to this conclusion, we have highlighted in the Figure the clusters for which the background is determined to better than 1\% of the cluster's central surface brightness (an uncertainty level below which the cluster profiles do not appear to be grossly disturbed by incorrect background subtraction, see Figure \ref{fig:profile background}). These high quality surface brightness profiles follow the mean trend. In the cases where the King model does significantly better than the EFF models and the cluster is a rich cluster with a well determined sky brightness and overall profile, the difference in the models is clearly in the outermost regions (see for examples clusters 85, 86, 96, 143, 174). In other cases, the cause for the preference of the King model is more subtle and ambiguous. It is presumably due to a cumulative effect in the fit over the range between core and tidal radius.

\begin{figure}[tbp]
\begin{center}
\plotone{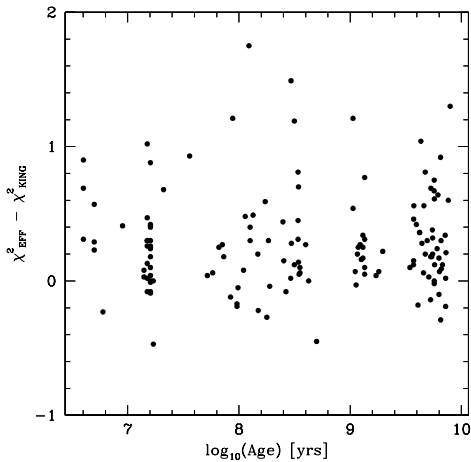}
\caption{The $\Delta \chi_\nu^2$ values vs. age. We include only clusters for which
either the King or EFF models fit with $\chi_\nu^2 < 1$ and $\Delta \chi^2_\nu < 2$.}
\label{fig:chi_age}
\end{center}
\end{figure}

To examine whether one particular model is favored for clusters of a particular age,
for example EFF models for the youngest clusters, we have plotted the difference
in $\chi_{\nu}^2$ values as a function of age (ages from \cite{rz}) in Figure \ref{fig:chi_age} for fits where
either the King or EFF model had $\chi_\nu^2 < 1$. There is no
trend evident that favors one particular model at a specific age. The general preference
for King models ($\Delta \chi_\nu^2 > 0$) is seen at all ages. Of course this result doesn't 
argue against the effectiveness of the EFF profile for some clusters, nor against the
existence of some clusters that are not tidally limited. Nevertheless, the situation is clearly more
complicated than an expectation that all young clusters should have power-law surface
brightness profiles.

\section{Discussion}

\subsection{Cluster Subpopulations}

\begin{figure}[tp]
\begin{center}
\plotone{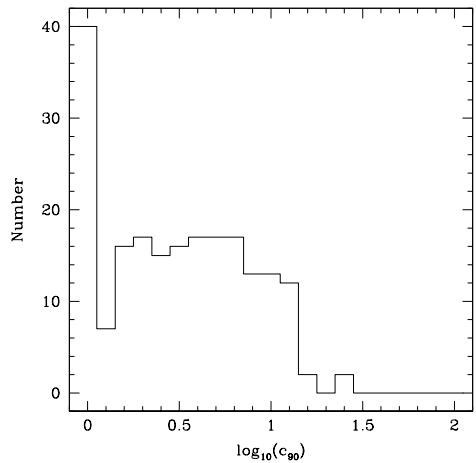}
\caption{The distribution of cluster concentrations in the SMC.  Note that $r_{90}$ as opposed to tidal radius is used to calculate the concentrations for this Figure, while the standard 
definition of concentration is fit in the model.  The axes are set so that the entire range of allowable values of concentration is shown.}
\label{fig:c90loghist}
\end{center}
\end{figure}

	Even those clusters that are well-fit by a King profile are not drawn from a single parameter population.  Although the core radius, $r_c$, and $r_{90}$ are correlated at the 92.7\% confidence level (Spearman rank test; see Table \ref{tab:Spearman Correlations}), clusters distribute themselves over the entire range of  $r_{90}/r_c$ that we allow in our fitting space given the theoretical constraints described by \cite{w85} (Figure \ref{fig:c90loghist}).  This implies that a variety of factors influence the observed structural parameters. Possible influences include tidal effects, age (the internal dynamical evolution of the cluster), and initial conditions (substructure, velocity anisotropies, initial stellar mass function, angular momentum, and chemical abundance). These are notoriously difficult to disentangle, but the large, uniform sample presented here offers a new opportunity to search for clues.

	We find King profiles to be a statistically acceptable fit to clusters that range over a factor of over 40 in $r_{c}$, 10 in $r_{90}$, and $10^{3}$ in $\Sigma_0$.  This finding is somewhat surprising in light of the fact that the vast majority of clusters are inferred to be in the process of dissolution or disruption on the basis of the cluster age distribution \citep{bl,rz} and many have had insufficient time to dynamically relax.  Therefore,
the fits to some clusters are almost certainly not physically meaningful, but rather are 
statistically allowed given the large observational uncertainties. Nevertheless, even in 
these cases our measurements do reflect the size of the cluster and its core. Although the large majority of SMC clusters are adequately fit by simple King profiles, and hence demonstrate
that measuring surface brightness profiles alone is not a promising avenue to identifying dissolving
clusters, there are several populations of clusters that merit further discussion and may be showing signs of their dynamical evolution.

\subsubsection{The Ring Clusters}

\begin{figure}[t]
\begin{center}
\plotone{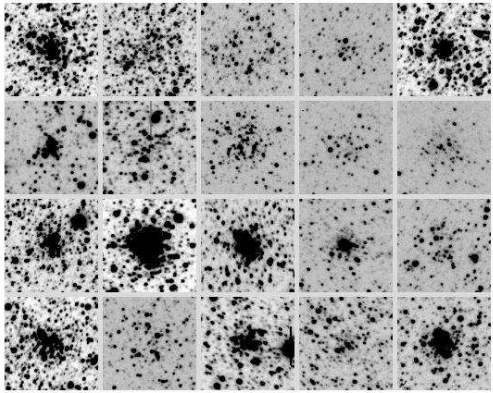}
\caption{{V-band images of the twenty clusters with the largest $\chi_\nu^2$'s. From
upper left across and then down these are clusters 71, 89, 19, 17, 61, 155, 88, 186, 39, 15, 70, 145, 87, 181, 114, 53, 148, 81, 77, and 143.}}
\label{fig:badclusters}
\end{center}
\end{figure}

	A subset of our cluster luminosity profiles contain a significant bump in the luminosity profile relative to the fitted King or EFF profile.  Of those, visual inspection suggests that about 12 out of the 204  have a bump that occurs beyond the central 8 arcsec and is not apparently due to a few luminous stars (clusters 12, 19, 39, 102, 115, 148, 164, 179, 183, 186, 196, and 203 in Figure \ref{fig:Sample Profiles}) . 
Although the apparent ring features in some of these clusters may be due to statistical fluctuations, 
there are at least some where the central deficit of stars is quite evident (see clusters 19, 39,
148, and 186 in Figure \ref{fig:badclusters}) and the $\chi_{\nu}^2$ values, for these clusters, 
are larger than expected for random fluctuations in a sample of this size with 99.5\% 
confidence for either the King or EFF profiles (see \S3.2).
We discount similar features at small radius, for example in clusters 3 and 4, because they could arise from slight centering offsets or random fluctuations within the small inner annuli. We also have not included some clusters in which the profile bump appears to be due to bright contaminating stars (such as cluster 82). Nevertheless, there may be many more clusters in the sample than the 12 listed in which the bump is a real feature.  One of the best examples of this phenomenon is cluster 19 (Figures \ref{fig:Sample Profiles}, \ref{fig:Sample Images}, and \ref{fig:badclusters}). 
This feature may be a signal of a cluster's dissolution because such a configuration appears
to be difficult to maintain in equilibrium. Further dynamical modeling is required to confirm this
supposition. Aside from dynamical issues, it is also the case that because we fit to the luminosity profile rather than the stellar density profile, we  are susceptible to variations in the mass (or number) to light ratio.  The degree to which the underlying light profile departs from a King or EFF profile may not indicate a similarly strong deviation in the mass profile.

\begin{figure}[t]
\begin{center}
\plotone{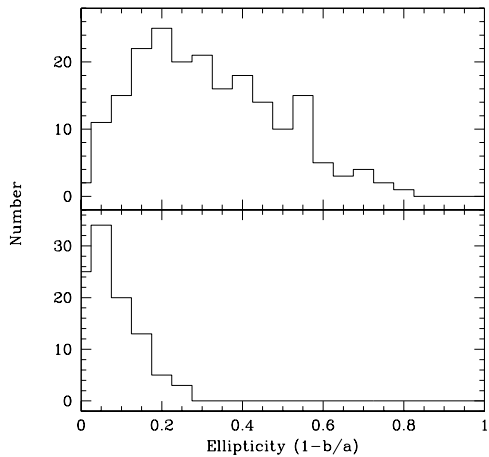}
\caption{The distribution of cluster ellipticities in the SMC and MW.  The SMC distribution is depicted in top panel while that of the MW appears in the bottom panel. The MW data come from \cite{Harris}}
\label{fig:e2mwvssmc}
\end{center}
\end{figure}

\subsubsection{The Ellliptical Clusters}\label{sec:ellip}

	The distribution of cluster ellipticities is presented in Figure \ref{fig:e2mwvssmc}.  To explore the origin of the apparent broad distribution relative to that of the Galactic clusters (data from
the compilation of \cite{harris}), we search for significant correlations between ellipticity and other cluster properties. For example, ellipticity has been found to correlate inversely with age \citep{FF}, which suggested either evolutionary changes in the structure of clusters or changes in the conditions of formation. In a different interpretation of the same correlation,  \cite{e91} suggested that subclumps at various stages of merging may produce the impression of large ellipticity in LMC clusters.  This conclusion was based on the 
inverse correlation between age and ellipticity and the interpretation of various features in cluster profiles.  However, \cite{van84} find that although ellipticity and age are indeed inversely correlated, that result is a byproduct of the inverse correlation between ellipticity and luminosity, a claim not addressed by \cite{e91}.  However, it is quite difficult to work in terms of luminosity since it depends both on mass
and age, both of which might help determine the ellipticity.

\begin{figure}[t]
\begin{center}
\plotone{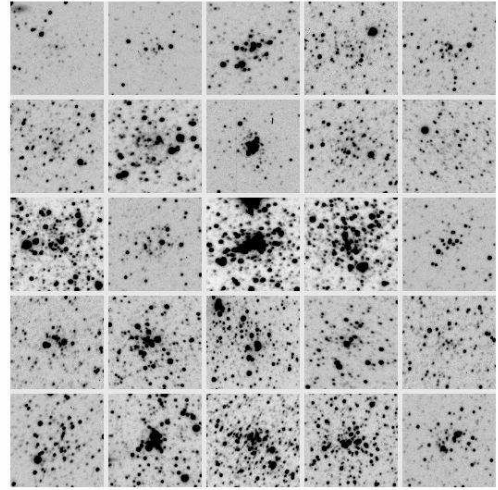}
\caption{{V-band images of the 25 most elliptical clusters in the sample}}
\label{fig: high_e}
\end{center}
\end{figure}

\begin{figure}[t]
\begin{center}
\plotone{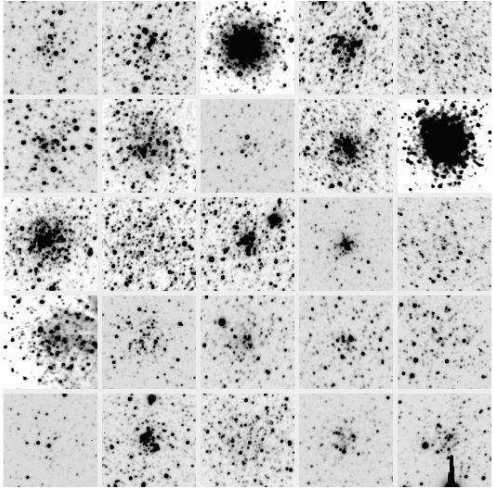}
\caption{{V-band images of the 25 least elliptical clusters in the sample}}
\label{fig: low_e}
\end{center}
\end{figure}

\begin{figure}[t]
\begin{center}
\plottwo{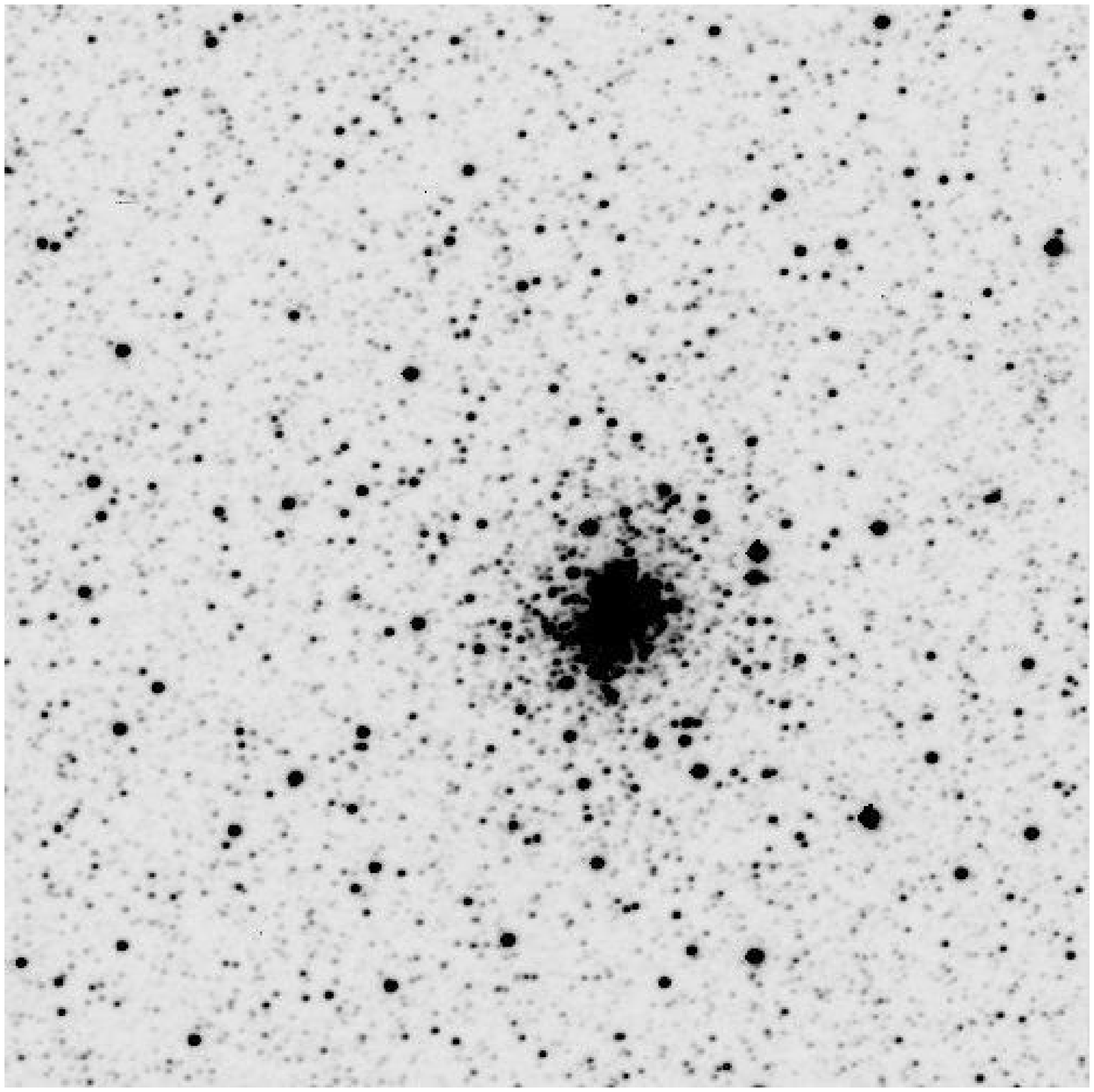}{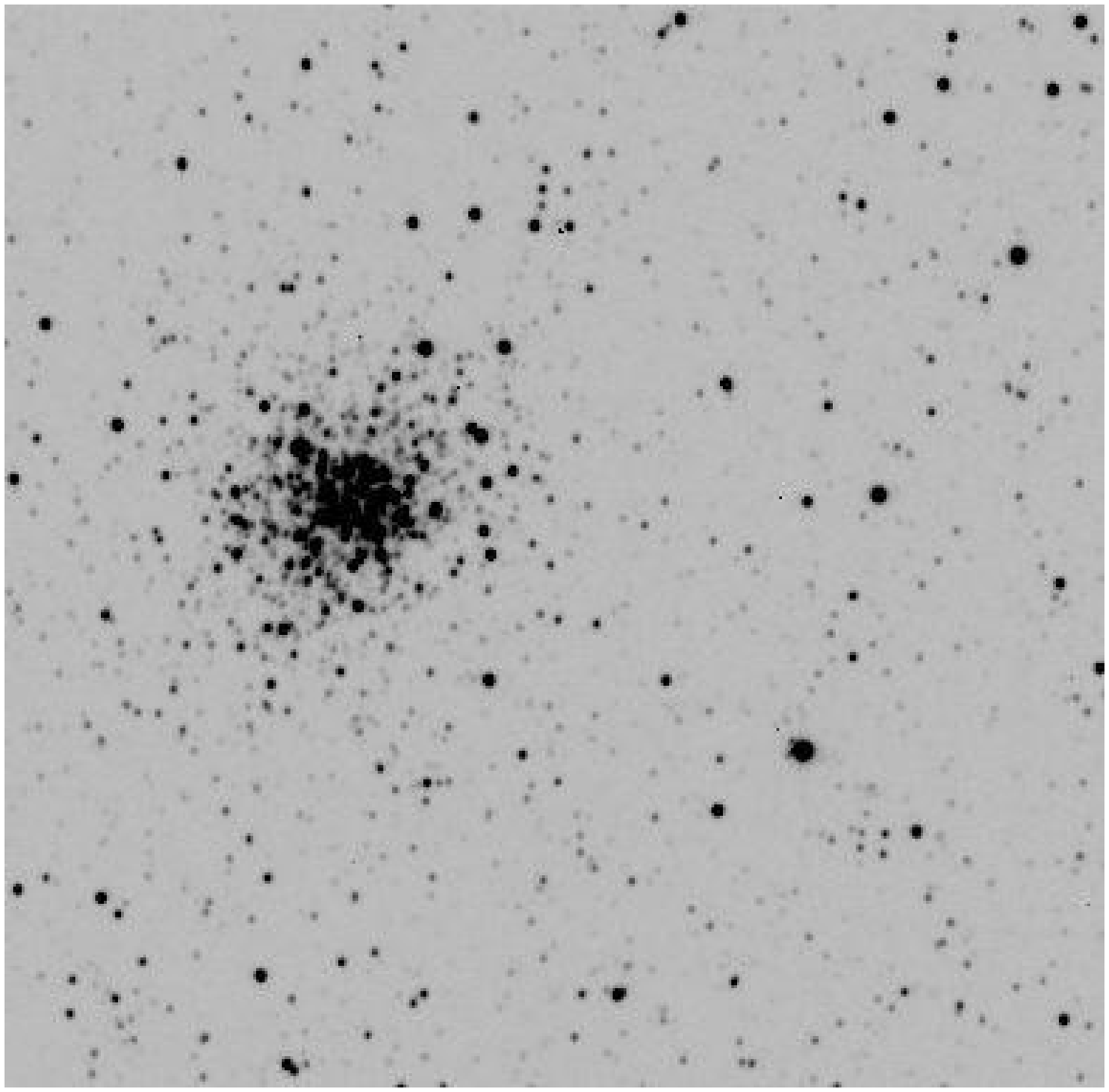}
\caption{{V-band images of cluster 98 and 197.}}
\label{fig: examples}
\end{center}
\end{figure}

	In Figures \ref{fig: high_e} and \ref{fig: low_e} we present the images of the 25 most and least elliptical clusters. The most elliptical clusters are in general also highly irregular, as well as of low luminosity. The least elliptical clusters are in general, although not exclusively, richer systems. While there are comparably poor, low luminosity systems that are both of high and low ellipticity, the luminous systems appear to uniformly avoid the upper tail of the ellipticity distribution. Two examples of rich clusters with moderately high ellipticities ($\sim 0.2$) are presented in Figure \ref{fig: examples}. These are among the most elliptical of the rich clusters in the SMC and interestingly correspond to the highest ellipticities  measured for Galactic clusters. 

\begin{figure}[t]
\begin{center}
\plotone{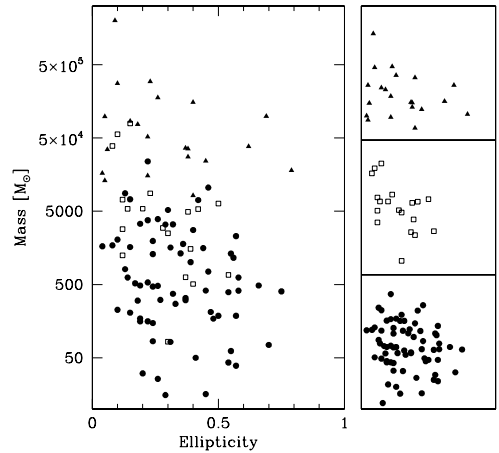}
\caption{{Cluster ellipticity vs. mass for three different age bins. The left panel
shows all the data, where the youngest clusters at represented with circles, the
intermediate age clusters with squares, and the oldest clusters with triangles (see
text for defining ages). The right
panels separate the three age populations for clarity and rescale to maximize
the plot for the relevant population.}}
\label{fig:erz1}
\end{center}
\end{figure}

We use the age dating results of \cite{rz} to examine whether age or mass is the primary
driver of ellipticity. We have converted the age and luminosity measurements from \cite{rz}
for the $Z=0.004$ models 
into a mass by using the same Starburst99 models \citep{s99} used to estimate the ages. In Figure \ref{fig:erz1}, we plot ellipticity vs. mass for different age bins.
The three bins are defined such that the youngest bin contains clusters whose 90\% upper
confidence age is less than 500 Myr, the middle bin contains clusters whose best
fit age is between 1 and 3 Gyr, and the oldest bins contains clusters whose 90\% lower
confidence age is greater than 3 Gyr.   The mean ellipticity of clusters in the
three age bins are 0.31, 0.27, and 0.32 (from youngest to oldest), and so we find no correlation between age and ellipticity. On the other hand, all three age bins show some sign
of a correlation with mass. All the data together results in a significant
detection of an inverse correlation (Spearman rank correlation confidence levels of
99.1\%). Results for the individual age bins suggest similar inverse
correlations, but the correlation
is only significant in the intermediate age bin (97.5\% confidence). Examining the 
joint panel illustrates that it is not possible to form large samples of clusters of similar
mass with which to examine trends with age. For example, all but seven of the young
clusters have masses less than 5000 $M_\odot$, while none of the older clusters
have such low masses. Selecting clusters by luminosity 
mixes lower mass young clusters with higher mass old clusters and so results
in a comparison of more massive old systems, with less massive young systems. 
While our results show that the dependence of ellipticity on mass is more easily
identified than any with age, there is a relative deficit of young clusters with
very low ellipticity, suggesting that age, and dynamical evolution, may still play a role.
	
\subsubsection{The Irregular Clusters}

	In Figure \ref{fig:badclusters} we show the images of the 20 clusters with the highest $\chi_{\nu}^2$ values relative to the King model fitting. These also have high $\chi_{\nu}^2$ values relative
to the EFF model fitting. Several of these clusters are the ``ring"-type clusters described above (19, 39, 148, 186), which explains why these are poorly fit by the either profile. Others have very high central luminosities (70, 71, 181).  The remainder appear to suffer from luminosity fluctuations caused by a few stars, which are perhaps unassociated with the cluster. Our principal conclusion is that with the exception of the ring clusters, we find no systematic failure mode for the King 
and EFF profile fitting. Even in cases where $\chi_\nu^2$ is large, the fits appear to be sufficiently good that one can ascribe the mathematical failure to such systematics as contamination. 

\subsection{Correlations and Comparisons}
	
	We present a set of Spearman rank correlation measures among the cluster parameters we considered (Table \ref{tab:Spearman Correlations}). Such an exercise can often lead to spurious correlations that are in fact driven by hidden variables, but it can also provide clues to fundamental physical drivers. We discuss a few of the statistically significant correlations that we find interesting. First, we find an inverse correlation between core radius and central surface brightness.  Such a correlation would appear to arise naturally among a set of clusters that are dissolving.  Second, we find a correlation between the central surface brightness and the background level. This could arise from the practical difficulty in identifying low-surface brightness clusters in denser environments, but could also be due to more efficient disruption of such clusters in
higher density regions \citep{go}.  These correlations are all internally consistent with a model that has
clusters preferentially disrupting in higher density regions.
One can only proceed so far on the basis of these correlations. Two key inputs are necessary to continue: 1) ages to disentangle evolutionary effects and 2) models that make quantitative predictions for these correlations. We are currently working on  joining the available age estimates \citep{rz} with
model predictions to address these questions.

\begin{deluxetable*}{crrrrrrrr}
\tablecaption{Spearman Rank Correlation Probabilities}
\tablewidth{0pt}

\tablehead{
\colhead{Parameter} &
\colhead{$r_{c}$} &
\colhead{$r_{90}$} &
\colhead{$c_{90}$} &
\colhead{Central Den.} &
\colhead{Luminosity} &
\colhead{Ellipticity} &
\colhead{Background} &
}

\startdata
$r_{c}$ & \nodata& $.927$& $-1$& $-1$& $1$& $.741$& $-.957$\\
$r_{90}$& $.927$& \nodata& $1$& $-.434$& $1$& $-.678$& $.369$\\
$c_{90}$ & $-1$& $1$& \nodata& $1$& $-.613$& $-.659$& $.951$\\
Central Den.& $-1$& $-.434$& $1$& \nodata& $1$& $-.974$&$.998$\\
Luminosity & $1$& $1$& $-.613$& $1$& \nodata&$-.991$& $.909$\\
Ellipticity& $.741$& $-.678$& $-.659$& $-.974$& $-.991$& \nodata& $-.874$\\
Background& $-.957$& $.369$& $.951$& $.998$& $.909$& $-.874$& \nodata\\
\enddata
\tablecomments{Probability of correlation between listed cluster properties computed with Spearman rank analysis.  Negative sign denotes anti-correlation.}
\label{tab:Spearman Correlations}
\end{deluxetable*}

\begin{figure}[h]
\begin{center}
\plotone{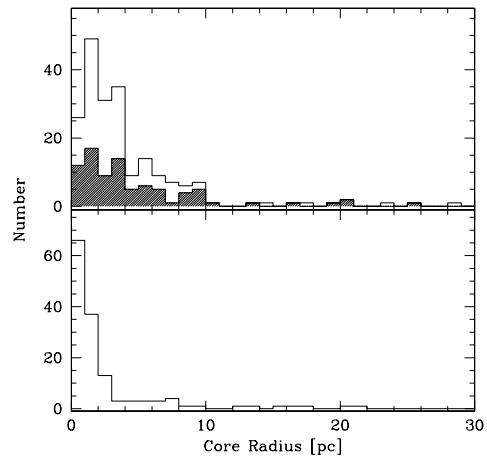}
\caption{The distribution of core radii in the SMC and MW.  The SMC distribution is depicted in top panel while that of the MW appears in the bottom panel. In the top panel we also show
the distribution for SMC clusters with age $>$ 1 Gyr in the shaded historgram. The MW data
come from \cite{Harris}}
\label{fig:coremwvssmc}
\end{center}
\end{figure}

	To complement the internal comparison approach, we also compare the cluster properties to those of Milky Way clusters (data for Galactic clusters from the compilation of \cite{harris}).  We present a comparison of the core radii of Milky Way and SMC clusters in Figure \ref{fig:coremwvssmc}.  The distribution of core radius size is much broader in the SMC than in the MW. 
This difference is either due to the preferred dissolution of young, low-concentration clusters
(both in the SMC and MW) and the preponderance of young clusters in the SMC, or to the
preferred formation and survival of lower-concentration clusters in the SMC.  The distribution
of core radii for clusters with age $>$ 1 Gyr (Figure \ref{fig:coremwvssmc}) shows that the older population is just as
broad in its distribution of core radii and demonstrates that long-lasting, low-concentration
clusters are more prevalent in the SMC than in the MW. Similar results were found for a smaller
SMC cluster sample, as well as cluster samples for the LMC, Fornax,  and Sagittarius dSphs 
\citep{mac04}.

\section{Summary}

	We present structural parameters for 204 stellar clusters in the Small Magellanic Cloud derived from fitting King and Elson, Fall, and Freeman (EFF) profiles to the V-band surface brightness profiles as measured from the Magellanic Clouds Photometric Survey images. We find that King profiles do a remarkable job at fitting the vast majority of the clusters, which is particularly surprising given that 99\% of the clusters will dissolve over the next few hundred million years \citep{rz} and that many have had insufficient time to dynamically
relax. We find that the King profiles systematically fit slightly better than the EFF
model for most clusters in our sample, independent of cluster age,
but that on a cluster by cluster basis it is 
difficult to rule out one type of model or the other. 
The only systematic deviation in the surface brightness profiles that we identify is in a set of clusters that lack a central concentration, and which we have named as ``ring" clusters.
In agreement with previous studies, we find that the clusters in the SMC are significantly more elliptical than those in the Milky Way. 
We are able to identify an inverse correlation between cluster mass and ellipticity, but
find no evidence for one between age and ellipticity. The search for the 
latter is hampered by the lack
of a sizeable sample of clusters with similar masses and a range of ages.
We identify several correlations (central surface brightness vs. local background density, size vs. distance) that an be used to constrain models of cluster evolution
Much of the interpretation of the empirical results we present rests on having measured ages and a detailed model for the formation and rapid evolution of the cluster population (extending the models of \cite{bekki}). This current sample of structural parameters provides the basis for detailed studies of the evolution of the cluster system of the Small Magellanic Cloud.

\acknowledgements
We gratefully acknowledge the referees' thorough and careful review of this manuscript.
AH acknowledges financial support from a NASA Spacegrant.  
DZ acknowledges financial support from National Science Foundation
CAREER grant AST-9733111, AST-0307482, and a fellowship from the David and Lucile
Packard Foundation.

\clearpage
\clearpage

\clearpage


\begin{thebibliography}{}

\bibitem[Banks et al.(1995)]{banks}
Banks, T., Doff, R. J., \& Sullivan, D. J. 1995, \mnras, 272, 821

\bibitem[Bekki et al.(2004)]{bekki}
Bekki, K., Beasley, M. A., Forbes, D.A., \& Couch, W. J. 2004, \apj, 602, 730

\bibitem[Bertin and Arnouts(1996)]{sextractor}
Bertin, E., \& Arnouts, S. 1996, \aaps, 117, 393

\bibitem[Bica and Dutra(2000)]{bica}
Bica, E., \& Dutra, C.M. 2000, \aj, 119, 1214

\bibitem[Bica et al.(1999)]{bicalmc}
Bica, E.L.D., Schmitt, H.R., Dutra, C.M., and Oliveira, H.L. 1999, \aj, 117, 238

\bibitem[Binney and Tremaine(1987)]{bt}
Binney, J., \& Tremaine, S. 1987, {\sl Galactic Dynamics}, (Princeton Univ. Press; Princeton)

\bibitem[Boutloukos \& Lamers(2003)]{bl}
Boutloukos, S.G., \& Lamers, H.J.G.L.M. 2003, \mnras, 338, 717

\bibitem[Da Costa and Freeman(1976)]{daf}
Da Costa, G. S., \& Freeman, K. C. 1976, \apj, 206, 128

\bibitem[Elson(1991)]{e91}
Elson, R.A.W. 1991, \apjs, 76, 185

\bibitem[Elson et al.(1987)]{e87}
Elson, R.A.W., Fall, S.M., \& Freeman, K.C. 1987, \apj, 323, 54

\bibitem[Elson et al.(1989)]{e89}
Elson, R.A.W., Freeman, K.C., \& Lauer, T.R. 1989, \apj, 347, L69

\bibitem[Frenk and Fall(1982)]{FF}
Frenk, C.S., \& Fall, S.M. 1982, \mnras, 199, 565

\bibitem[Goodwin(1997)]{goodwin}
Goodwin, S. P. 1997, \mnras, 286, L39

\bibitem[Gnedin \& Ostriker(1997)]{go}
Gnedin, O. Y., \& Ostriker, J. P. 1997, \apj, 474, 223

\bibitem[Grillmair et al.(1995)]{grill}
Grillmair, C. J., Freeman, K.C., Irwin, M. \& Quinn, P.J. 1995, \aj, 109, 2553

\bibitem[Gunn and Griffin(1979)]{gg}
Gunn, J.E., \& Griffin, R.F. 1979, \aj, 84, 752

\bibitem[Han \& Ryden(1994)]{hr}
Han, C., \& Ryden, B. S. 1994, \apj, 433, 80

\bibitem[Harris(1996)]{harris}
Harris, W. E. 1996, \aj, 112, 1487

\bibitem[Harris \& Zaritsky(1999)]{hz}
Harris, J. \& Zaritsky, D. 1999, \aj, 117, 2831

\bibitem[Holland, Cot\'e, \& Hesser(1999)]{holland}
Holland, S., Cot\'e, P., \& Hesser, J. E. 1999, \aa, 348, 418

\bibitem[Holtzman et al.(1992)]{holtzman92}Holtzman, J.A. et al.
1992, \aj, 103, 691

\bibitem[King(1962)]{k62}
King, I. 1962, \aj, 67, 471

\bibitem[King(1966)]{k66}
King, I. 1966, \aj, 71, 64

\bibitem[Kontizas et al.(1982)]{kon82}
Kontizas, M., Danezis, E., \& Kontizas, E. 1982, \aaps, 49, 1

\bibitem[Kontizas \& Kontizas(1983)]{kon83}
Kontizas, E. \& Kontizas, M. 1983, \aaps, 52, 143

\bibitem[Kontizas et al.(1986)]{kon86}
Kontizas, M., Theodossiou, E., \& Kontizas, E. 1986, \aaps, 65, 207

\bibitem[Leitherer et al.(1999)]{s99}
Leitherer et al. 1999, \apjs, 123, 3

\bibitem[Lupton et al.(1989)]{lup}
Lupton, R.H., Fall, S.M., Freeman, K.C., and Elson, R.A.W. 1989, \apj, 347, 201L

\bibitem[Lupton \& Gunn(1987)]{lg}
Lupton, R. H., \& Gunn, J. E. 1987, \aj, 93, 11106

\bibitem[Lupton et al.(1987)]{lgg}
Lupton, R. H., Gunn, J. E., \& Griffin, R. F. 1987, \aj, 93, 1114

\bibitem[Mackey \& Gilmore(2003a)]{mac03a}
Mackey, A.D., \& Gilmore, G.F. 2003a, \mnras, 338, 85

\bibitem[Mackey \& Gilmore(2003b)]{mac03b}
Mackey, A.D., \& Gilmore, G.F. 2003b, \mnras, 338, 120

\bibitem[Mackey \& Gilmore(2004)]{mac04}
Mackey, A.D., \& Gilmore, G.F. 2004, \mnras, 355, 504

\bibitem[Michie(1963)]{michie}
Michie, R. W. 1963, \mnras, 125, 127

\bibitem[Pietrzy\'nski et al.(1998)]{ogle}
Pietrzy\'nski, G., Udalski, A., Kubiak, M., Szyma\,nski, M., Wo\'zniak, P. \& Zebru\'n, K.
1998, Acta Astron., 48, 175

\bibitem[Prendergast \& Tomer(1970)]{pt}
Prendergast, K.H., \& Tomer, E. 1970, \aj, 80,427 

\bibitem[Rafelski \& Zaritsky(2004)]{rz}
Rafelski, M., \& Zaritsky, D. 2004, \aj, submitted

\bibitem[van den Bergh(1981)]{van81}
van den Bergh, S. 1981, \aaps, 46, 79

\bibitem[van den Bergh and Morbey(1984)]{van84}
van den Bergh, S., \& Morbey, C.L. 1984, \apj, 283, 598

\bibitem[Webbink(1985)]{webbink}
Webbink, R. F. 1985, in IAU Symp. 113 Dynamics of Star Clusters, ed. J. Goodman \& P. Hut  (Dordrecht:Reidel),  541

\bibitem[Whitmore et al.(1993)]{whitmore}Whitmore, B.C., Schweizer, F., Leitherer, C., Borne, K.,
\& Robert, C. 1993, \aj, 106, 1354

\bibitem[Whitmore et al.(1999)]{whit}
Whitmore, B.C., Zhang, Q., Leitherer, C., Fall S.M., Schweizer, F., Miller, B.W. 1999, \aj, 118, 1551

\bibitem[Wilson(1975)]{wilson}
Wilson, C. P. 1975, \aj 80, 175

\bibitem[Wiyanto et al.(1985)]{w85}
Wiyanto, P., Kato, S., \& Inagaki, S. 1985, \pasj, 37, 715

\bibitem[Yang et al.(2004)]{yang}
Yang, Y., Zabludoff, A. I., Zaritsky, D., Lauer, T. R., \& Mihos, J. C. 2004, \apj, 608, 358

\bibitem[Zaritsky et al.(1997)]{z97}
Zaritsky, D., Harris, J., \& Thompson, I. 1997, \aj, 114, 1002

\bibitem[Zaritsky et al.(2002)]{z02}
Zaritsky, D., Harris, J., Thompson, I., Grebel, E.K., \& Massey, P. 2002, \apjs, 

\bibitem[Zaritsky et al.(1996)]{z96}
Zaritsky, D., Shectman, S.A., \& Bredthauer, G 1996, \pasp, 108, 104


\end{thebibliography}
\end{document}